\documentclass[sigconf]{acmart}
\usepackage{amsmath,amsfonts}
\usepackage{algorithmic}
\usepackage{graphicx}
\usepackage{textcomp}
\usepackage{xcolor}
\usepackage{multirow}
\usepackage{multicol}
\usepackage{booktabs, tabularx} 
\usepackage{subcaption}
\usepackage{float}
\usepackage[export]{adjustbox}
\usepackage{scalerel}

\AtBeginDocument{%
  }

\setcopyright{acmlicensed}
\copyrightyear{2024}
\acmYear{2018}
\acmDOI{XXXXXXX.XXXXXXX}

\acmConference[ESEM'24]
{18th ACM/IEEE International Symposium on Empirical Software Engineering and Measurement}
  {September 03--05,
  2024}{Barcelona, Spain}
\acmISBN{978-1-4503-XXXX-X/18/06}

\newenvironment{myquote}[1]%
{\list{}{\leftmargin=#1\rightmargin=#1}\item[]}%
{\endlist}

\begin{document}
\title{Usefulness of data flow diagrams and large language models for security threat validation: a registered report}
\pagestyle{plain}

\author{Winnie Bahati Mbaka}
\email{w.mbaka@vu.nl}
\orcid{0000-0001-6913-1971}
\affiliation{%
  \institution{Vrije Universiteit Amsterdam}
  \country{The Netherlands}
}

\author{Katja Tuma}
\email{k.tuma@vu.nl}
\orcid{0000-0001-7189-2817}
\affiliation{%
  \institution{Vrije Universiteit Amsterdam}
  \country{The Netherlands}}

\renewcommand{\shortauthors}{Mbaka et al.}
\newcommand{\comt}[2][]{\colorbox{purple}{{\scriptsize\bfseries\color{white}COMMENT}} {\color{purple}#2} \colorbox{purple}{\scriptsize\color{white}#1}}
\newcommand{\commentKT}[1]{\comt[KT]{#1}}
\newcommand{\new}[1]{{\color{black} #1}} 

\begin{abstract}
  The arrival of recent cybersecurity standards has raised the bar for security assessments in organizations, but existing techniques don't always scale well. 
  Threat analysis and risk assessment are used to identify security threats for new or refactored systems. Still, there is a lack of definition-of-done, so identified threats have to be validated which slows down the analysis.
  Existing literature has focused on the overall performance of threat analysis, but no previous work has investigated how deep must the analysts dig into the material before they can effectively validate the identified security threats.
  We propose a controlled experiment with practitioners to investigate whether some analysis material (like LLM-generated advice) is better than none and whether more material (the system's data flow diagram and LLM-generated advice) is better than some material.
  In addition, we present key findings from running a pilot with 41 MSc students, which are used to improve the study design.
  Finally, we also provide \new{an initial} replication package, including experimental material and data analysis scripts \new{and a plan to extend it to include new materials based on the final data collection campaign with practitioners (e.g., pre-screening questions).}
\end{abstract}





\maketitle

\section{Introduction}\label{sec:intro}
Building secure software is a global concern that has in recent years spurred new regulations (e.g., EU Cybersecurity and Cyberresilience Acts, and the US Cloud Act). CISA and 17 U.S. and international partners~\cite{cisaSBD} recommend planning for countermeasures and reduce the risk of costly security breaches down the line, and for safety-critical systems, such as systems developed in the transportation sector, recent standards even require conducting a threat analysis of entire products (ISO/SAE 21434:2021~\cite{isosae21434}).
\new{In light of the global gap in the cybersecurity workforce, these demands can be disruptive for organizations.}

Threat analysis and risk assessment approaches help to elicit critical security threats and identify appropriate countermeasures. 
\new{A myriad of threat analysis and risk assessment techniques already exists \cite{tuma2018threat}.
 STRIDE \cite{shostack2014threat} adopts a graphical representation of the software architecture under analysis, the \textit{Data Flow~Diagram} (DFD) \cite{deng2011privacy,sion2018solution}, which is simple and easy to learn~\cite{scandariato2015descriptive}. 
    DFD-like models (in essence, directed typed graphs) are extensively used in practice, for instance, in the automotive industry~\cite{macher2016review}, at Microsoft~\cite{shostack2014threat}, and agile organizations~\cite{bernsmed2019threat}.
}
\new{In a threat analysis session, security and domain experts explore the DFD (and the available material)} to identify
potential security threats
through brainstorming~\cite{shostack2014threat}.
\new{But this process is time-consuming~\cite{scandariato2015descriptive}, and the resources dedicated to security are scarce in organizations.
In addition, threat analysis lacks completeness guarantees about the identified security threats~\cite{tuma2018threat,cruzes2018challenges}.
}
\new{Without enough objective measures for threat correctness and completeness~\cite{mbaka2023measures},}
predicting threats for software that has not been built yet demands practitioners to make risk-based decisions under uncertainty~\cite{Bier2020}. 
In other words, practitioners \new{often waste time revisiting the analysis material and \textit{validating threat feasibility}~\cite{tuma2021finding}} to assert that they have done a "complete job" and that they have not overlooked any important security threat.
%
%
For example, practitioners may quickly assess \new{the threat of SQL-injection on an exposed web interface} as correct by reading the threat description and attack scenario, while for \new{more domain-specific threats}, they may have to revisit analysis material (such as the DFD, the requirements of the system, general or domain-specific security catalogues, such as MITRE ATT\&CK knowledge base\footnote{https://attack.mitre.org}).

\new{
Previous attempts to automate threat analysis have limitations \cite{tuma2018threat}, as the underlying system under analysis is under-specified at design-time.
}
Large Language Models (LLMs) have \new{recently} been studied for providing security advice and summarizing information~\cite{chen2023can}.
As an alternative to manual use of general security catalogs, \new{practitioners could} adopt LLMs as Model-in-the-loop~\cite{faggioli2024determines}, where the human decides based on summarized advice of the LLM.
\new{In fact, tools for leveraging advice from LLMs for STRIDE are emerging \cite{STRIDE-GPT-talk2024,STRIDE-TM-mentor2024}.
For example, a customized GPT~\cite{STRIDE-TM-mentor2024} provides suggestions for prompts, which generate a list of potential security threats. 
Even if in-house models are deployed in a secure environment,
LLMs tend to hallucinate~\cite{huang2023survey},
so the generated list of threats may be eventually used as \textit{additional} analysis material.
%
%
}
\smallbreak
\textit{So, how deep must the analyst dig into the material before they can effectively validate security threats?}

We plan to investigate the usefulness of analysis material for threat validation, with the goal of understanding if reading "some" additional analysis material is better than "none" and if "more" available analysis material is better than "some". 
\new{
Since practitioners tend to refer to (and revisit) the DFD during threat validation by assessing threat priority and feasibility~\cite{tuma2021finding}, it is interesting to include the presence of a DFD for this task as an intervention.
In addition, having the DFD available could potentially help the practitioners in assessing the LLM advice, therefore it is also interesting to observe the presence of both DFD and LLM advice.
}

\new{
Due to known issues with hallucinations~\cite{huang2023survey}, we expect that if participants trust the first generated advice, this can increase the number of false positives.
On the other hand, we expect that participants not fully trusting the first generated advice may perform better.
Our results could inform practitioners about the minimal analysis material required for effective threat validation, and provide insights to practitioners that are on the fence about adopting LLMs for threat analysis.
}

We present a balanced design for a controlled experiment (approved by the Ethics Board of the primary institution conducting the research), a pilot study with 41 MSc students, and a plan for execution with practitioners.
We built a ground truth using the Microsoft STRIDE~\cite{shostack2014threat} threat analysis approach on two scenarios (GitHub repository update, and pod deployment on K8), and 
since the final intervention is ordinal, we plan to conduct a statistical analysis using Helmert contrasts.

\subsection*{Data availability statement}
\new{For replicability, we have provided an initial replication package\footnote{https://anonymous.4open.science/r/LLM-for-threat-validation-298B/README.md} 
including a sample analysis script,
the entry questionnaire,
the experiment survey, both scenario descriptions, and a list of threats for each scenario. We plan to extend it in the final paper to include pre-screening questions for recruiting industry practitioners.}







\section{Related Works}
\label{sec:related}

\subsection{Empirical Research of Threat Analysis}

Several studies \cite{scandariato2015descriptive, mbaka2023measures} have empirically investigated the effectiveness of Microsoft's threat model, STRIDE. Both Scandariato et al \cite{scandariato2015descriptive} and Mbaka et al. \cite{mbaka2023measures} replicated a controlled experiments that sought to measure the productivity, precision, and recall of STRIDE. 

Mbaka et al. \cite{mbaka2023measures} compared the productivity and precision
of two STRIDE techniques, per-element and per-interaction. 
In the case of Mbaka et al. \cite{mbaka2023measures} STRIDE-per-interaction teams performed better than their counterparts. However, 
no significant location shift between the productivity and precision of the two variants was reported.

Other studies have investigated the challenges of adopting STRIDE in agile software development ~\cite{cruzes2018challenges,bernsmed2019threat}. 
Bernsmed et al. \cite{bernsmed2019threat} conducted interviews with employees from four organizations that use agile software development practices 
and observed several challenges for adoption.
Yet, practitioners were in agreement that performing a threat analysis leads to a more secure product. 


Apart from STRIDE, other threat analysis techniques, have been empirically investigated. Two studies have compared the effectiveness of attack trees and misuse cases \cite{karpati2014comparing, opdahl2009experimental}. 
Karpati et al.~\cite{karpati2014comparing} observed that attack trees resulted in the identification of a higher number of threats compared to misuse cases, albeit the difference was not statistically significant. Although the experimental set-up implemented by Opdahl et al. \cite{opdahl2009experimental} was different from that of Karpati et al. \cite{karpati2014comparing} (student participants vs industry practitioners) similar observations of better performance when using attack trees were also reported in the former~\cite{opdahl2009experimental}.

Diallo et al. \cite{diallo2006comparative} compared the applicability of three techniques (i.e., the common criteria, attack trees, and misuse case) applied to wireless hotspots. 
The authors observed that each technique has strengths and weaknesses. For instance, while common criteria were complex to learn, they were easy to analyze. On the other hand,  misuse cases had easier learnability, but its output was not easy to read.

\vspace{0.7\baselineskip}
\noindent
\fbox{\begin{minipage}{0.96\columnwidth}
Similar to~\cite{scandariato2015descriptive, mbaka2023measures} we adopt the STRIDE methodology to measure the correctness of validating identified security threats. In contrast, instead of tasking the participants with developing a DFD, which is challenging~\cite{mbaka2023measures}, we present them with the same DFD in the experimental material. In addition, none of the empirical studies on threat analysis \cite{scandariato2015descriptive, mbaka2023measures,karpati2014comparing, opdahl2009experimental,diallo2006comparative} have investigated how to support human reasoning around security threat validation.
\end{minipage}}

\subsection{Application of LLMs in Security}
Related work has evaluated the usefulness of LLMs in the detection of code and software vulnerabilities \cite{szabo2023new,thapa2022transformer,omar2023detecting,sun2023gpt,cheshkov2023evaluation}. 
Omar et al. \cite{omar2023detecting} proposes a vulnerability detection framework, VulDetect leveraging the strengths of three models (GPT-2, BERT, and LSTM) to detect C and C++ source code vulnerabilities.
The authors observed that VulDetect was able to outperform other state-of-the-art vulnerability detection techniques (SyseVR and VulDeBert) with a 92.65\% accuracy.

Cheshkov et al. \cite{cheshkov2023evaluation} sought to investigate the effectiveness of large language models (ChatGPT and GPT-3) in vulnerability detection. 


In the study by Sun and colleagues \cite{sun2023gpt} GPTScan (a combination of GPT and static analysis) was proposed and used to detect logic vulnerabilities in smart contracts. 

Szabo and Bilicki \cite{szabo2023new} evaluated the efficacy of the GPT-3.5 and GPT-4 models in detecting the CWE-653 weakness, identifying
sensitive data, and determining the protection levels of front-end applications. 

Chen and colleagues \cite{chen2023can} investigated the ability of two LLMs (ChatGPT and Bard) to refute popular security and privacy misconceptions. 
The study reported that LLMs report false positives (supporting misconceptions) which increase when the misconceptions are repeatedly queried. The study also reported on LLM hallucination, providing invalid URLs to support their output~\cite{chen2023can}.




\vspace{0.7\baselineskip}
\noindent
\fbox{\begin{minipage}{0.96\columnwidth}
Despite some benefits of LLMs in discovering and assessing security vulnerabilities
\cite{chen2023can,szabo2023new,thapa2022transformer,omar2023detecting,sun2023gpt,cheshkov2023evaluation}, 
previous research has not investigated their use for early security design analyses, such as finding security weaknesses~\cite{szabo2023new}, and no previous work
investigated their use for security threat validation.

\end{minipage}}
\section{Research Questions}
Our research is motivated by previous findings pointing to the challenges with threat analysis reproducibility~\cite{sion2020security,mbaka2023measures} and lack of definition-of-done~\cite{cruzes2018challenges, galvez2018odyssey}. 
A significant challenge is determining the feasibility of the identified threats, or in other words, validating the identified threats, an activity that can take place several times during the process and has been measured to cause detours and slow down the progress~\cite{tuma2021finding,iqbal2020requirement}.
Threat validation is not only an issue at the design phase, but also during threat intelligence gathering~\cite{islam2022smartvalidator}, where the collected data is also tainted by uncertainty.

On the other hand, previous research 
has
investigated the effectiveness of providing additional textual or graphical material to aid in the comprehension of functional requirements \cite{abrahao2012assessing} and safety compliance needs \cite{de2020empirical}.
We postulate that 
understanding what material (if any) should be used to effectively validate threats
may help in deriving threat feasibility faster.

To this end, we formulate the first research question:

\textbf{RQ1: What is the actual usefulness of having additional material like DFD or LLMs during threat validation?}

To measure the actual usefulness of having additional analysis materials (DFDs, LLMs, or both), we define several treatment groups. First, we consider the performance of participants validating threats without any additional material. To this end, we first compare their performance to those who received some (a DFD, or an LLM).
Second, we compare the performance of those who received some material to those who received both a DFD and LLM.
We hypothesize that actual usefulness should be different between groups that did not receive additional materials to those that did. 
That is, having some additional material (a DFD, or LLM) raises the performance of participants. 
In addition, having both a DFD and LMM 
should
increase the participant’s ability to correctly identify realistic
threats.
%
We therefore propose the following alternative hypothesis;


$H_{performance}$: \textit{There is a statistically significant difference in the actual usefulness (i) between participants assessing the validity of threats without additional materials to those with some (either a DFD, or an LLM) and (ii) between participants assessing the validity of threats with some additional materials (DFD or LMM) to those with both (DFD and LLM).}

Second, some previous research measured differences~\cite{liu2021graphical} in the perceived usefulness of graphical models compared to textual information for security risk assessment.
But, Labunets et al. \cite{labunets2017equivalence} found that
tabular and graphical methods are statistically equivalent to each other with respect to the actual and perceived efficacy.
%
Following up on this result, we also measure perceived usefulness of the material provided to support threat validation and expect to confirm equivalence in our study.


\textbf{RQ2: What is the perceived usefulness of the additional material during threat validation?}

To this end, we check for statistical equivalence using the alternative hypothesis formulated below; 

$H_{equiv-perc-both}$: \textit{When given both the DFD and LLM, their perceived usefulness is statistically equivalent.}


Second, the sample means of the treatment group that received only the DFD will be compared to the one asked to assess the correctness of threats using only LLM. We formulate an alternative hypotheses;

$H_{equiv-perc-isolation}$: \textit{The perceived usefulness of DFDs and LLMs when used in isolation is statistically equivalent.}


\section{Recruitment and Ethical Concerns}
\subsection{\new{Pilot study}}
\label{subsec:participants}
We define two target populations for this study. First, for the pilot study, we recruited Master Computer Science students attending courses taught by the experimenters. 
To ensure that the students understand the experimental objects necessary to conduct the study we
conducted a \new{4h30min} training, evaluated their understanding of the training material, and included attention checks to filter dishonest responses.

\subsection{\new{Participant Recruitment}} \label{subsec: recruitment}
\new{We plan to conduct the study with professionals recruited from crowd-sourcing platforms (e.g., Upwork, or Prolific).} To this end, we will \new{pre-screen and recruit} participants with a background in software development, cybersecurity risk management. 
Since threat modeling is carried out by domain experts, such as developers, software architects, and security specialists~\cite{cruzes2018challenges}, we consider such sample as representative.

\new{In addition, we are conducting a think-aloud protocol with a few recruited practitioners as a second, more qualitative pilot. The results of the think-aloud will be transcribed and analysed for potential issues and concerns raised by the practitioners that need to be addressed before the final data collection.}

\subsection{Ethical Concerns}
This experiment has received ethical approval from the ethical board of the institution under review number 2024-013. 
First, the study will provide an opt-in consent ("yes"/"no") form. 
Second, we do not anticipate any potential risk to the participants or researchers. Third, the study utilises GDPR-compliant tools to collect data. In addition, any personally identifiable information will be removed before data analysis. 
Fourth, for the pilot, the study was conducted as part of a course, and participants were only incentivised with a participation point which has a very small effect on their final grade. Importantly, students who did not provide consent for the analysis of their data also received a participation point. Finally, the participants were debriefed on the process of the experiment and the artefacts used.

\section{Methodology}
This section presents the design of the experiment
including the research plan.

\subsection{Experimental design}
\label{subsec:design}
Assigning each participant to a different condition (noLLM, LLM) x (noDFD, DFD) x (GH, K8) would require $(2)^3$ groups. To avoid such a huge number of groups we will make use of a balanced orthogonal design which is also known as Taguchi Design~\cite{kacker1991taguchi}.
Each participant will be randomly assigned to one of the four groups:

\begin{table}
    \centering
    \footnotesize{}
    \begin{tabular}{cccc}
    \toprule
     &\multicolumn{3}{c}{Task ($\times$ 2)} \\
        Groups &  DFD& LLM & Scenario\\
         \midrule
         Group A & \checkmark & \checkmark & GH,K8\\
         Group B & \checkmark & - & GH,K8\\
         Group C & - & \checkmark & GH,K8\\
         Group D & - & - & GH,K8\\
         \bottomrule
    \end{tabular}
    \caption{Full experimental design used in the pilot. }
    \label{tab:design}
\end{table}
\begin{enumerate}
 
\item  LLM + DFD (A) receives the scenario descriptions with an accompanying data flow diagram instance and tasked with assessing the applicability of threats using an LLM
\item noLLM + DFD (B) receives the scenario descriptions with an accompanying data flow diagram instance and tasked with self-assessing the applicability of threats
\item LLM + noDFD (C) receives the scenario description without an accompanying data flow diagram instance and tasked with assessing the applicability of threats using an LLM
\item noLLM + noDFD (D) receives the scenario description without an accompanying data flow diagram instance and tasked with self-assessing the applicability of threats
\end{enumerate}

\subsection{Experimental Objects}\label{subsec:objects}
To ensure the objectives of the study are met, we prepared several experimental objects.

\paragraph{Scenario selection} 
First, we run the study with two scenarios to increase generalizability.
One is based on updating a remote repository on GitHub and one on deploying a pod on Kubernetes, two common tasks in software development.
In addition, both scenarios are inspired by real open-source platforms.



\paragraph{LLM selection} Several Large Language Models have been developed including Google's Gemini, OpenAI's chatGPT, GitHub's Copilot, and Microsoft's Copilot among others. Since the aim of our study is not to train an LLM or to benchmark different LLMs, but rather investigate their usefulness for helping human analysts in validating threats, we plan to leverage an open-source model. To this end, we opted for chatGPT-3.5 turbo model, as a first step. 
%

\paragraph{Training \new{(4h30min)}} \new{For the pilot}, the experimenters have prepared \new{a lecture on threat modeling process and landscape (2h), a training lecture with a deep dive into security threats, STRIDE, DFD, and scenarios\footnote{The training delivered to groups A and B also included a DFD of the scenario.} (2h), and a walk-through presentation (30min) detailing each step of the task. The training was delivered in the beginning of the week on two consecutive days, and was also made available as recordings.}
The walk-through offers more understanding of what is to be expected during the actual experiment. For instance, groups A and C's walk-through includes an example of prompting the LLM to assess the correctness of the threat.



\paragraph{Ground Truth}
\new{The ground truth has been developed systematically by the authors. 
Half of the threats presented to the participant are correct and feasible, and half are bogus or incorrect. 
} 
We define a bogus threat as the creation of a false claim to the existence of a potential security risk.     
Below is an example of a bogus threat:
\begin{myquote}{0.1in}
\textit{Scenario:} Updating a remote repository on GitHub \\
\noindent
\textit{Threat description:} An unauthenticated and non privileged attacker can still submit custom code into the remote repository to prepare the first step of another attack (e.g., turning off logging service or cause a Denial of Service).\\
\noindent
\textit{Assumption:} The attacker can reach the remote repository  (e.g. through internet).\\
\noindent
\textit{STRIDE category:} Elevation of privilege and Tampering\\
\noindent
\textit{Location (in DFD):} The remote code repository    
\end{myquote}

Explanation: GitHub allows owners of repositories to specify branch protection rules, which essentially disables 'force push' to the matching branches and prevents the matching branches from being deleted. When branch protection rules are implemented, an attacker cannot submit a custom code.

\paragraph{Measures of Success}

Table~\ref{tab:vars} presents all the variables we consider in our experiment. This study considers two independent variables (or intervention), i.e., providing a DFD as part of the hand-out material and asking participants to perform the task of the experiment using an LLM. To achieve the aims of this study, we first plan to measure the background experience of participants in relation to the experimental objects of the study. In this case, participants will be required to self-report on their prior experience with secure design techniques, software design models, and their usage of LLMs, GitHub, and cloud deployment platforms. The responses to their background experience will be captured either on a 5-point Likert scale or using predefined multiple-choice options.

Second, 
to answer our first research question, we will analyse participants' performance 
, against the ground truth. 
The four possible outcomes of performance are discussed below; 

\begin{enumerate}
    \item True Positive (TP), correctly identified realistic threats 
    \item True Negative (TN), correctly identified bogus threats 
    \item False Positive (FP), bogus threats selected as being real
    \item False Negative (FN), real threats that were considered bogus and therefore not selected

\end{enumerate}

Third, we measure the perceived usefulness of graphical models and LLMs in threat validation. To this end, participants will be asked to what extent, on a 5-point Likert scale\footnote{Where point 1 is labeled "strongly disagree", point 3 is labeled "neutral", and point 5 is labeled "strongly agree".}, do they think the additional analysis materials were useful in assisting them to correctly identify realistic threats. The responses to this question will be used to answer our second research question.

Lastly, we include several control questions to account for the varying levels of comprehension of the experimental objects among the participants. These control measures consist of questions about participants understandability of the the task, the sufficiency of time allocated to complete the task alongside the sufficiency of the training materials. In addition, we include several attention and background checks for each of our target populations (see subsection \ref{subsec:participants}). For student participants, we will measure their understanability of the experimental objects via attention checks. For industry practitioners, we plan to have several checks to ensure that they have a technical background. 
\new{Finally, we collect the prompt history for each participant to control for prompt variability.}

\begin{table*}[]
\footnotesize
    \centering
    \caption{Experimental Variables}
    \label{tab:vars}
    \begin{tabular}{p{0.2\textwidth} p{0.48\textwidth} p{0.12\textwidth}}
        \toprule
         \textbf{Name} & \textbf{Description} & \textbf{Operationalization} \\
         \midrule
         \multicolumn{3}{l}{\textit{Independent variables (design)}} \\
         \midrule
        
         DFD &  Receiving a DFD to support the threat validation & Nominal (*) \\
         LLM & Receiving LLM API to support the threat validation & Nominal (*) \\
        \midrule
         \multicolumn{3}{l}{\textit{Background experience variables}} \\
         \midrule
         Secure design & Self-reported experience with secure design techniques  & Ordinal scale ($\dagger$) \\
         Modeling & Self-reported familiarity with design models & Ordinal scale ($\ddagger$) \\
         Use of LLM & Self-reported frequency and the reason for using LLMs  & Ordinal scale  ($\dagger$)\\
         Scenario 1 & Self-reported experience with GitHub & Ordinal scale  ($\dagger$)\\
         Scenario 2 & Self-reported experience with cloud deployment platforms  & Ordinal scale  ($\dagger$)\\
         \midrule
         \multicolumn{3}{l}{\textit{Dependent variables}} \\
         \midrule
         Performance of threat validation & Participants performance evaluated against a ground truth & Interval scale ($\equiv$) \\
         Perception of additional materials &
         Self-reported perceived usefulness of additional analysis materials & Ordinal scale  ($\dagger$)\\
         
         \midrule
         \multicolumn{3}{l}{\textit{Control measures}} \\
         \midrule
         Understanding & Self-reported understanding of what the task required them to do & Ordinal scale ($\ddagger$)\\
         Time & Self-reported sufficiency of the allotted time & Ordinal scale ($\ddagger$) \\
         Training & Self-reported sufficiency of training & Ordinal scale ($\ddagger$) \\
         Attention checks & To account for participants understanding of the experimental objects &  Interval scale ($\equiv$)\\
         Background checks (practitioners) & To account for the professional background of industry practitioners recruited from crowd-sourcing platforms & Interval scale ($\equiv$)\\
        \bottomrule
    \end{tabular}
    \begin{minipage}{0.9\textwidth}
\vspace{0.5\baselineskip}
\begin{description}
    \item[(*)] Qualtrics configured to automatically randomise the allocation of the independent variables.
    \item[($\dagger$)] Multiple choice: For experience - attended some lecture, attended a full course, short internship, professional engagement. \\ 
    For frequency- never, several times a day, once a week, few times a month, once in 3 months, or "other-asked to specify".
    \item[($\ddagger$)] Responses captured on a 5-point Likert scale.
    \item [($\equiv$)] Responses evaluated against a predefined ground truth
\end{description}
\end{minipage}
    \end{table*}

\section{Pilots}
\subsection{Execution} \label{subsec:execution}
We present the steps taken to execute the pilot study.

Each participant \textit{p} joining the experiment;
\begin{enumerate}
    \item was randomly assigned to one of the four treatment groups: A, B, C, and D, see subsection \ref{subsec:design}.
    \item was presented with two scenario descriptions, one on modifying and updating repositories on GitHub (GH) and the other on pod deployment on Kubernetes (K8). 
        We configured the survey tool\footnote{www.qualtrics.com/} such that the presentation of the scenarios is randomised. That is, for two participants in the same groups, one received the Kubernetes scenario first followed by the GitHub scenario, and vice versa for the second participant. 
    \item was presented with a list of threats (five bogus and five realistic threats) to each scenario description.

    \item was tasked with assessing the correctness of each threat and select the threats considered as realistic (likely to occur).
    \end{enumerate}
Each threat was accompanied by a threat description, assumptions, the associated STRIDE threat category that would be compromised if the attack occurred, and affected components. For each threat marked as realistic, participants were required to provide a short justification as to why they think it is a real threat.  

\new{When assessing the validity of security threats using LLMs (groups A and C), participants were allowed the freedom to prompt the LLM as they would in a real-world scenario. We provided them with the body of the prompt (threat description, assumption, STRIDE threat type, and affected components) to be posted on the LLM and asked them to prompt (in their own words) the LLM for assistance. We recorded the entire LLM interaction for each security threat selected as realistic by participant. 
} 
    

\subsection{\new{Preliminary} results}\label{subsec:results}
\emph{Demographics.} In total, 41 participants joined the \new{pilot} study, \new{each participant received both scenarios, so we collected in total 82 responses.} 
Before the training, about $2/3$ reported to have used GitHub in a professional capacity (13) or during an internship (12), and more than $2/3$ 
reported having attended either a few lectures (18) or a full course on Kubernetes (10). 
About half reported (22) to be new to the topic of secure design and most of the participants in groups A and C reported
using an LLM several times a day. 

\emph{\new{Results}.} Groups that were tasked with validating security threats with the help of an LLM performed slightly better 
(the means for correctly identifying security threats was slightly higher in groups A ($\mu$TP$\_$A= 8.1) and C ($\mu$TP$\_$C= 9.4) compared to groups B ($\mu$TP$\_$B= 6.0) and D ($\mu$TP$\_$D= 7.4)). \new{However, higher false positives were also reported in groups with access to an LLM (A ($\mu$FP$\_$A= 7.4) and C ($\mu$FP$\_$C= 4.7) compared to groups B ($\mu$FP$\_$B= 3.5) and D ($\mu$FP$\_$D= 3.1)). Similar observations (incorrectly assessing security information) have also been reported in prior studies on security misconceptions \cite{chen2023can}. Chen and colleagues \cite{chen2023can} reported that LLMs incorrectly support popular security and privacy misconceptions.}
\emph{\new{Preliminary} observations.} 
\new{
In addition to the descriptive statistics, we inspected the responses manually and made some interesting observations. However, the sample in the pilot is relatively small and these observations need to be validated with practitioners in the final study.

Interestingly, the group with no additional material (D) reported on average the least number of FPs but still reported a relatively high number of correct threats (7.4 out of 10). If this finding is validated with practitioners, this would indicate that less analysis material is actually needed to validate threats.

We investigated the responses of participants who received LLM advice (A, C) and observed that all participants in these groups took the first advice generated by LLM and also reported lower levels of previous knowledge about security (10 were novice to security, 9 attended some lectures, and only 2 had some hands-on experience). 
This could potentially also explain the higher numbers of FPs due to LLM hallucinations in those groups.
In addition, when asked about the perceived usefulness of LLMs in assessing the validity of security threats after each scenario. We observed that 10/21 participants strongly agreed or agreed (points 5 and 4 on the Likert scale) with LLM's perceived usefulness concerning the threats relating to the GitHub scenario. On the other hand, 14/21 participants strongly agreed or agreed 
with LLM's perceived usefulness concerning the threats relating to the Kubernetes scenario. 
Thus, for security novices (e.g., practitioners in training), the use of LLM advice may still help in terms of recall (not overlooking threats) but not precision (due to a higher chance of false positives).

We also found that} the presence of DFD did not make a significant difference in the collected measures of TN, FP, FN. Further, no major difference was observed for TPs when comparing the groups C and D to A and B (t-test of $\alpha$=0.05 returned a p-value of 0.048 and the Pearsons' correlation statistic was 0.5). We conclude that no strong positive or negative correlation exists between the availability of a DFD and the ability to correctly identify realistic threats. 

%
This observation indicates that, instead of the full factorial design from Table~\ref{tab:design}, it is important to observe the following progression: performance differences when no material is given vs when LLM is used vs when DFD is given and LLM is used.

\section{Data Analysis Plan}
\label{sec:data analysis}
Once the full data collection with practitioners has been completed, the results will be aggregated and statistically analysed. 

\emph{RQ1.} 
We aim to analyse the data using the Helmert contrast, a statistical analysis used to determine the smallest shift in location of the intervention from the control treatment \cite{ruberg1989contrasts}. We formulate the problem for difference as;



\begin{enumerate}
    \item noDFD\&noLLM vs (noDFD\&LLM union DFD\&LLM) 
    \item noDFD\&LLM vs  DFD\&LLM 
\end{enumerate}

To this end, if noDFD\&LLM union DFD\&LLM is greater than the control group (noDFD\&noLLM), then \textit{"some"} additional analysis material may improve the effectiveness of threat validation than having no materials at all. Similarly, if DFD\&LLM is greater than noDFD\&LLM, then both graphical models and LLMs may increase the actual effectiveness of threat validation as opposed to only having access to one of the additional analysis materials. 

\emph{RQ2.} Since our data for measuring perception of usefulness is ordinal and may not be normally distributed, we plan to use Mann Whitney U (MWU) with a level of significance equal to 0.05  ($\alpha =0.05$) to test both equivalence and difference. We formulate the problem for testing the statistical equivalence as;

\begin{eqnarray*}
p_{low}  & =  & MWU (\{x - \delta | x \in A\}, B, alt = 'less') \\
p_{up} & = & MWU (B,\{x + \delta | x \in A\}, alt = 'less')
\end{eqnarray*}

Where $A$ and $B$ are the vectors of the dependent variables (perceived usefulness of DFD and LLM). The $\delta$ represents the range for which we consider the means of both groups to be equivalent. The value of $\delta$ will be determined before the analysis. While estimating the value of delta might seem arbitrary, similar approaches have been used in Food and Drug surveys~\cite{meyners2012equivalence}. 

\section{Threats to Validity} \label{sec:threats}
This section discusses the planned mitigations to potential internal and external threats to validity.

\paragraph{Internal validity} 
\new{
We removed the security threat used in the walk-through video from the experiment task to avoid the risk of influencing participant responses.

To ensure that the complexity of the two scenarios is comparable, the DFDs have the same graph topology. Namely, both DFDs contain 3 data store nodes, 1 external entity node, 6 process nodes, and 16 data flows connecting the same type of nodes.
}

We also consider the threat of introducing experimenter bias in the ground truth. 
\new{
To mitigate this threat, we built the material carefully involving four researchers.
The DFD and list of security threats were built by two experimenters, one with more than 8 years of experience in threat
analysis, and one with more than 4 years of experience with Kubernetes and cloud security.
The ground truth was verified with two other group members (junior, and senior with $10+$ years experience in controlled experimentation).}

\new{
From the control measures, 29/41 participants reported either strongly agreeing or agreeing (points 5 and 4 on the Likert scale) to have a good understanding of what the task required them to do, and 30/41 either strongly agreed or agreed that the time allowed for the experiment was sufficient. To this end, we conclude that the training material and the time allowed to finish the task were sufficient for our target population.
}



\paragraph{External validity} 
\new{We are aware of the challenges in recruiting participants from crowd-sourcing platforms \cite{reid2022software,rauf2022challenges,ebert2022recruiting}, such as, participants self-reporting on their levels of expertise or background knowledge without having to provide evidence.
To mitigate this challenge, we will pre-screen the participants as recommended by Alami and colleagues \cite{alami2024you}. The pre-screening layer will include a set of questions 
on their self-reported knowledge and skills such as questions about the semantics of pseudocode for developers, questions about cybersecurity risk management practices, and questions about software architecture patterns. 
The pre-screening questions will be used to filter untrustworthy respondents.
}

We consider the threat of generalizability of our findings to real-world scenarios. 
\new{
To partially mitigate this risk, we use application of K8 pod deployment and Github remote repository update, two realistic and common tasks in software development, and allow participants to use LLMs in a controlled but not limiting way.
}

\new{While students participating in empirical studies have been reported to have a good understanding of industry-level requirements \cite{svahnberg2008using}, 
we decided to invite students in the first pilot study, but
plan to conduct the study with practitioners.
}

\section{Conclusion and Future work}
This paper presents the research and execution plan to conduct a study to measure the actual and perceived usefulness of graphical and Large Language Models in validating security threats. To this end, we intend to use a balanced orthogonal design with two interventions, DFD and LLM. 
\new{As a first step,} we ran a pilot with 41 students and \new{outlined the plans to carry out a think-aloud study before the final data collection with industry practitioners.} 

\section*{Acknowledgements}
This work was funded by the \textit{Nederlandse Organisatie voor Wetenschappelijk Onderzoek (NWO)} under the HEWSTI Project under grant no. 14261 and partly supported by the European Union under grant no. 101120393 (Sec4AI4Sec). 

\subsection*{CRediT statements}
	\emph{Conceptualization:} WM, KT; 
	\emph{Methodology:} WM, KT; 	
	\emph{Software:} NA ; 
	\emph{Validation:} WM, KT;	
    \emph{Formal analysis:} WM;	
    \emph{Investigation:} WM, KT;	
    \emph{Resources:}NA; 
    \emph{Data Curation:} WM; 	
    \emph{Writing - Original Draft:} WM, KT;
    \emph{Writing - Review \& Editing:} WM, KT; 
    \emph{Visualization:} WM; 
    \emph{Supervision:} KT; 
    \emph{Project administration:} KT; 
    \emph{Funding acquisition:} KT; 
\bibliographystyle{ACM-Reference-Format}
\bibliography{sample-base}

\end{document}